\documentclass[aps,prb,preprint,floatfix,superscriptaddress,showpacs]{revtex4-1}

\usepackage{graphicx}
\usepackage{dcolumn}
\usepackage{amsmath,bm}
\usepackage{mathtools}
\usepackage{color}%
\usepackage{multirow}
\usepackage{array} 
\usepackage{xr}
\usepackage{braket}
\usepackage{mdframed}
\usepackage{soul,xcolor}

\makeatletter
\newcommand*{\addFileDependency}[1]{
  \typeout{(#1)}
  \@addtofilelist{#1}
  \IfFileExists{#1}{}{\typeout{No file #1.}}
}
\makeatother
 
\newcommand*{\myexternaldocument}[1]{%
    \externaldocument{#1}%
    \addFileDependency{#1.tex}%
    \addFileDependency{#1.aux}%
}

\myexternaldocument{si}

\begin{document}
\setstcolor{red} 

\title{Generalized energy band alignment model for van der Waals heterostructures with a charge spillage dipole}

\author{Seungjun Lee}
\email{seunglee@umn.edu}
\affiliation{Department of Electrical and Computer Engineering, University of Minnesota, Minneapolis, Minnesota 55455, USA}
\author{Eng Hock Lee}
\affiliation{Department of Electrical and Computer Engineering, University of Minnesota, Minneapolis, Minnesota 55455, USA}

\author{Young-Kyun Kwon}
\affiliation{Department of Physics, Department of Information Display, and Research Institute for Basic Sciences, Kyung Hee University, Seoul, 02447, Korea}

\author{Steven J. Koester}
\affiliation{Department of Electrical and Computer Engineering, University of Minnesota, Minneapolis, Minnesota 55455, USA}

\author{Phaedon Avouris}
\affiliation{IBM T. J. Watson Research Center, Yorktown Heights, New York 10598, USA}
\author{Vladimir Cherkassky}
\affiliation{Department of Electrical and Computer Engineering, University of Minnesota, Minneapolis, Minnesota 55455, USA}
\author{Jerry Tersoff}
\affiliation{IBM T. J. Watson Research Center, Yorktown Heights, New York 10598, USA}

\author{Tony Low}
\email{tlow@umn.edu}
\affiliation{Department of Electrical and Computer Engineering, University of Minnesota, Minneapolis, Minnesota 55455, USA}
\affiliation{Department of Physics, University of Minnesota, Minneapolis, Minnesota 55455, USA}

\date{\today}
\begin{abstract}
The energy band alignment at the interface of van der Waals heterostructures (vdWHs) is a key design parameter for next-generation electronic and optoelectronic devices. 
Although the Anderson and midgap models have been widely adopted for bulk semiconductor heterostructures, they exhibit severe limitations when applied to vdWHs, particularly for type-III systems.
Based on first-principles calculations for approximately $10^3$ vdWHs, we demonstrate these traditional models miss a critical dipole arising from interlayer charge spillage. 
We introduce a generalized linear response (gLR) model that includes this dipole through a quantum capacitance term while remaining analytically compact.
With only two readily computed inputs, the charge neutrality level offset and the sum of the isolated-layer bandgaps, the gLR reproduces DFT band line-ups with $r^2\sim$~0.9 across type-I, II, and III stacks. 
Machine-learning feature analysis confirms that these two descriptors dominate the underlying physics, indicating the model is near-minimal and broadly transferable. The gLR framework therefore provides both mechanistic insight and a fast, accurate surrogate for high-throughput screening of the vast vdW heterostructure design space.
\end{abstract}

\maketitle


\section{Introduction}

The energy band alignment (EBA) of the heterostructure is the most important design parameter in modern electronic and optoelectronic devices. Since the 1970s, significant efforts have been devoted to understanding the EBA in bulk semiconductor heterostructures. These efforts led to the development of simple and physically intuitive EBA models such as Anderson, midgap, and related models.~\cite{anderson1988experiments,anderson1960germanium,harrison1988elementary,tersoff1984theory,tersoff1984theory2,tersoff1986band,van1989band,van1987theoretical,frensley1977theory,tejedor1978simple,langer1985deep,alferov1998history}
A common strength of these approaches is that they translate readily measurable or computable electronic properties of the individual materials (electron affinity, ionization potential, dielectric response, etc.) into a first-pass prediction of the composite EBA. Thus, these EBA models are invaluable for rapid materials screening and device prototyping: they supply reasonably accurate guidance at negligible computational cost and have steered the successful design of countless III–V and II–VI heterojunctions.

With the discovery of graphene and other two-dimensional (2D) materials, 
van der Waals heterostructures (vdWHs) have garnered significant attention over the last two decades.~\cite{geim2013van,Liu2016,science.aac9439,Song2024} 
Early work understandably emphasized the more prevalent type-I and type-II vdWHs, which underpin a wide range of optical and electronic devices.~\cite{Deng2014,Lin2015,Furchi2018,Liu2017,Binder2017} 
More recently, type-III vdWHs have attracted much attention because their broken gap configuration can be utilized in a multitude of electronic and photonic devices, including tunneling field-effect transistors,~\cite{Gong2013,Sarkar2015,Roy2015,szabo2015ab,Oliva2020,pal2024ultra} infrared photodetectors~\cite{tan2021broken,wang2023high} and other applications.
Unfortunately, the neglect of interfacial charge spillage and the quantum capacitance limits in existing models means that no compact, first-principles-informed model yet provides reliable predictions for type-III energy alignment in these atomically thin systems.

The enormous chemical diversity of the 2D material landscape provides an equally pressing incentive for a compact yet predictive EBA framework. 
For example, a recent 2D material database contains over 15,000 stable monolayers,~\cite{Haastrup_2018,Gjerding_2021} and thus the total number of available vdWHs exceeds $\sim$10$^8$.
Even with cutting‑edge computing resources, high‑throughput first‑principles screening of this vast materials space is intractable. 
Consequently, a simple, accurate, and physically intuitive EBA model will be significantly useful not only for reducing experimental and computational costs but also for deepening our understanding of the fundamental physics governing EBA in vdWHs.

To build and test a truly predictive EBA model for vdWHs, we concentrate on transition-metal dichalcogenides (TMDs), the best-studied 2D family and one that spans a wide spectrum of bandgaps, electron affinities, and ionization energies. 
Using 45 TMD monolayers, we constructed a total of 990 vdWHs and investigated their EBA through first-principles calculations based on density functional theory (DFT).
Our results clearly show that classical EBA models, which solely rely on one physical input feature - vacuum level or charge neutrality level (CNL) difference - fail dramatically in describing the EBA of a wide range of vdWHs.
We therefore introduce a generalized linear response (gLR) model that augments the CNL difference with the interfacial dipole produced by charge spillage. Using only two readily obtainable descriptors, the CNL difference and the sum of the isolated-layer bandgaps, the gLR reproduces the DFT EBAs with great accuracy. Independent machine learning feature importance analysis confirms that these very same descriptors dominate the physics, indicating that the gLR strikes an almost optimal balance between simplicity and accuracy.

\section{Results and Discussion}

\begin{figure*}[h]
\includegraphics[width=1.00\textwidth]{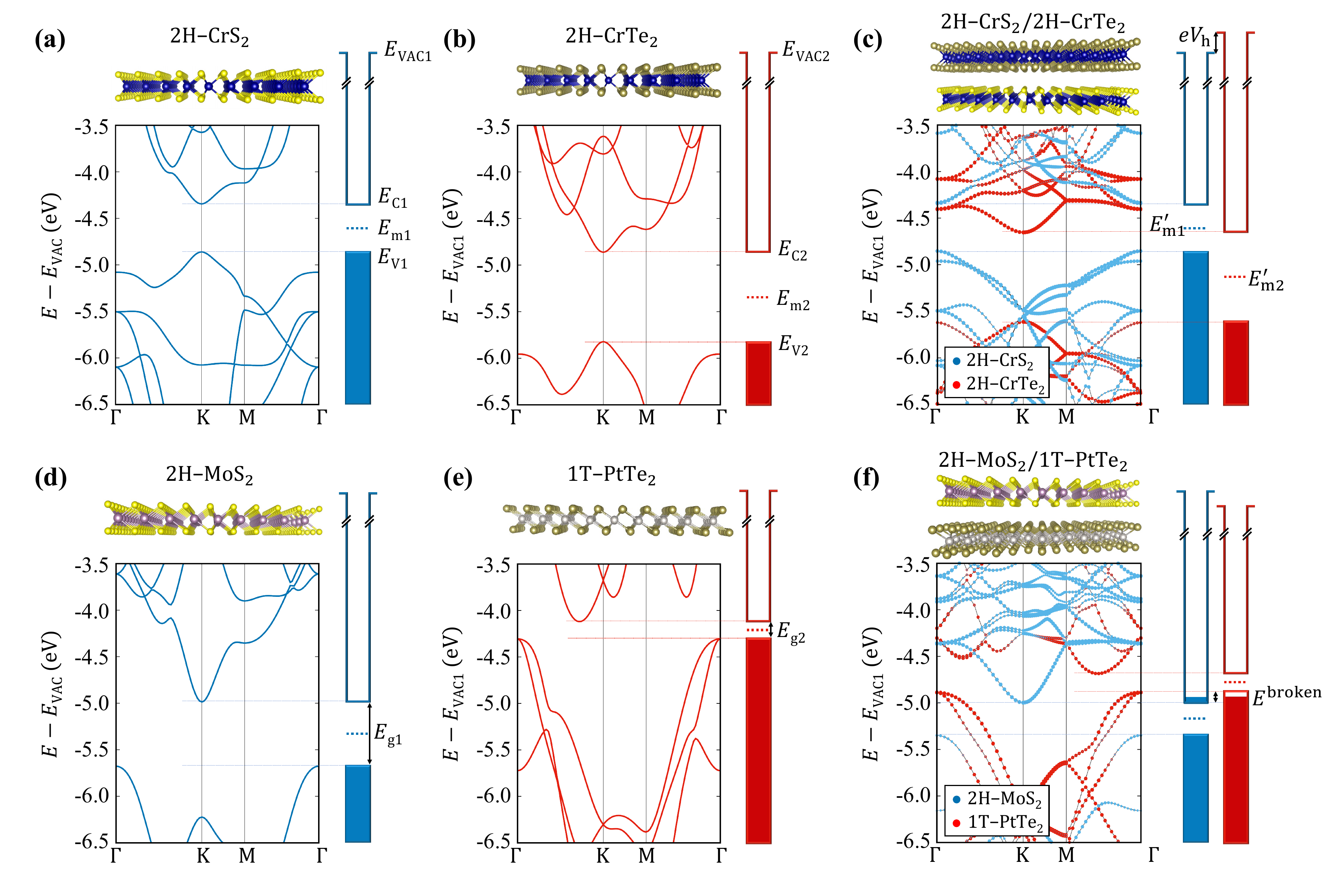}
\caption{
\textbf{Representative electronic structures of type-II and type-III vdWHs and their isolated monolayers.}  
Electronic structures and schematic energy band diagrams of isolated (a) 2H-CrS$_2$ and (b) 2H-CrTe$_2$ monolayers, respectively, and (c) their heterostructure.  
Electronic structures and schematic energy band diagrams of isolated (d) 2H-MoS$_2$ and (e) 1T-PtTe$_2$ monolayers, respectively, and (f) their heterostructure.  
In (c, f), the color coordinates represent the contribution of each layer, and the zero energy level corresponds to the vacuum level of layer 1.
}\label{fig1}
\end{figure*}
\subsection*{Energy level alignment and simple linear response model in van der Waals heterostructure}

The EBA in vdWHs is different from bulk heterostructures in several important ways.
Additionally, the weak vdW interaction at the interface typically does not significantly alter the electronic structures of the constituent monolayers once they form a heterostructure.
As a result, the bands of each monolayer can be regarded as rigid blocks that simply translate in energy when the layers are brought together. Secondary effects such as residual atomic relaxation or hybridisation introduce only minor corrections.

We test this “rigid-shift” picture on two instructive bilayers that bracket the regimes of interest with representative type-II and type-III vdWHs: the 2H-CrS$_2$/2H-CrTe$_2$ and 2H-MoS$_2$/1T-PtTe$_2$. We calculated the electronic structures of both the isolated monolayers and their heterostructure, as shown in Fig.~\ref{fig1}(a-f).
DFT calculations for the isolated monolayers and for the assembled stacks allow us to track four reference energies: the conduction band minimum ($E_{\rm{C}}$), valence band maximum ($E_{\rm{V}}$), and their averaged value, midgap energy ($E_{\rm{m}}$), and vaccum level ($E_{\rm{VAC}}$). 
The subscripts ``1'' and ``2'' refer to the two monolayers.
For the isolated cases, their $E_{\rm{VAC}}$ can be assumed to be zero energy. Within this choice of reference, for example, $E_{\rm{V}}$ of 2H-CrS$_2$ almost touches $E_{\rm{C}}$ of 2H-CrTe$_2$, and $E_{\rm{m1}}-E_{\rm{m2}}$ was calculated to be 0.746~eV.
When they form heterostructure, the wavefunction overlap between two layers generates a dipole step ($eV_{\rm{h}}$), as described in Fig.~\ref{fig1}(c).
By definition of the dipole step $eV_{\rm{h}}$, the midgap energy difference before and after interlayer interaction ($E_{\rm{m}}$ and $E'_{\rm{m}}$) can be expressed as 
\begin{equation}
(E'_{\rm{m2}}-E'_{\rm{m1}})=(E_{\rm{m2}}-E_{\rm{m1}})-eV_{\rm{h}},
\label{eq1}
\end{equation}
Using electrostatic potential distribution calculated by DFT, we found $eV_{\rm{h}}$=0.225~eV for 2H-CrS$_2$/2H-CrTe$_2$ heterostructure.
Layer-resolved band structures in Fig.~\ref{fig1}(c) also visualize that the bandgaps of 2H-CrS$_2$ and 2H-CrTe$_2$ remain nearly unchanged when they form heterostructure.
Compared to the isolated 2H-CrTe$_2$, $E_{\rm{C}}$ of 2H-CrTe$_2$ in vdWHs (with respected to $E_{\rm{VAC}}$ of 2H-CrS$_2$) are vertically shifted to 0.209~eV which is close to $eV_{\rm{h}}$.
This excellent agreement indicates that the $eV_{\rm{h}}$ primarily determines the EBA of vdWHs.

All of the foregoing arguments carry over to broken gap (type-III) stacks, as shown in Figure~\ref{fig1}(d-f). In type-III vdWHs, $E_{\rm{V}}$ of one layer sits above the $E_{\rm{C}}$ of the other layer, and this displacement is commonly called the broken gap ($E^{\rm{broken}}$). When we assume ${E'_{\rm{m2}}}>{E'_{\rm{m1}}}$, the positively-defined $E^{\rm{broken}}$ is described as, 
\begin{equation}
E^{\rm{broken}}
=E_{\rm{V2}}-E_{\rm{C1}}=(E'_{\rm{m2}}-E_{\rm{g2}}/2)-(E'_{\rm{m1}}+E_{\rm{g1}}/2)
=(E'_{\rm{m2}}-E'_{\rm{m1}})-\frac{1}{2}E_{\rm{g}}^{\rm{sum}}, 
\label{eq2}
\end{equation}
where $E_{\rm{g}}^{\rm{sum}}=E_{\rm{g1}}+E_{\rm{g2}}$ and $E_{\rm{g1}}$ and $E_{\rm{g2}}$ are bandgaps of the two constituent monolayers.
Because this overlap drives charge spillage from the higher-energy valence states into the lower-energy conduction states, it directly sets the interfacial dipole, and therefore the potential step $eV_{\rm{h}}$.
Thus, a practically useful EBA model must accurately predict $eV_{\rm{h}}$ for both type-II and type-III vdWHs.

To understand the interfacial dipole step $eV_{\rm{h}}$, we begin with the simple linear response (sLR) picture that has proven useful for bulk, organic, and vdW junctions.~\cite{tersoff1984theory,vazquez2005energy,PhysRevB.103.205129} 
The model rests on two physically transparent assumptions: First, the CNL of each monolayer is described by its midgap energy.~\cite{tersoff1984theory2,mulliken1934new,mulliken1935electronic}
Second, the dipole step grows linearly with the CNL offset between the layers, as follows~\cite{tersoff1984theory,vazquez2005energy,PhysRevB.103.205129} 
\begin{equation}
eV_{\rm{h}}={\alpha}_{\rm{eff}}(E'_{\rm{m2}}-E'_{\rm{m1}})
\label{eq3},
\end{equation}
where ${\alpha}_{\rm{eff}}$ is a dimensionless constant akin to the polarizability of vdWHs, though with different physical dimensions.
In this context, we refer to ${\alpha}_{\rm{eff}}$ as an ``effective'' polarizability, which arises from charge redistribution and wavefunction overlap between the two layers - quantum dipoles - at the interface.~\cite{tersoff1984theory2}
By solving Eq.~(\ref{eq1}) and Eq.~(\ref{eq3}) self-consistently, we arrive at the sLR model~\cite{PhysRevB.103.205129}
\begin{equation}
eV_{\rm{h}}=\frac{{\alpha}_{\rm{eff}}}{1+{\alpha}_{\rm{eff}}}(E_{\rm{m2}}-E_{\rm{m1}}).
\label{eq4}
\end{equation}
Eq.~(\ref{eq4}) can be further simplified by introducing a screening parameter of $S$ as follow~\cite{vazquez2005energy} 
\begin{equation}
eV_{\rm{h}}=(1-S)(E_{\rm{m2}}-E_{\rm{m1}}).
\label{eq5}
\end{equation}
$S$ directly measures how strongly the initial CNL difference is screened in the final alignment, which is more apparent when written as follows.
\begin{equation}
(E'_{\rm{m2}}-E'_{\rm{m1}})=S(E_{\rm{m2}}-E_{\rm{m1}}).
\label{eq6}
\end{equation}
It is worth noting that the well-known Anderson~\cite{anderson1988experiments,anderson1960germanium} and midgap models~\cite{tejedor1978simple,tersoff1984theory} are limiting cases of the sLR model corresponding to $S=1$ (no interfacial screening, ${\alpha}_{\rm{eff}}=0$) and $S=0$ (complete screening, ${\alpha}_{\rm{eff}}=\infty$), respectively.~\cite{PhysRevB.103.205129}

\subsection*{High-throughput first-principles calculations for energy level alignment of van der Waals heterostructure}

To map the EBA landscape systematically, we performed high-throughput DFT calculations on a total of 990 commensurate TMD bilayers.
Our monolayers were comprised of 20 2H- and 25 1T-phases of TMDs, which have negative formation energy and non-zero bandgaps, as predicted by the PBE calculation.~\cite{Rasmussen2015}
Commensurate supercells were generated with the coincidence-lattice approach that allows a small twist angle while limiting residual biaxial strain to $<$ 5\% (average $\approx$ 2\%).
To isolate interlayer interaction and more accurately investigate EBA, we need to compare monolayers and heterostructures with the same lattice constants.~\cite{PhysRevB.103.205129}
Therefore, we also calculated the electronic structure of each layer in isolation under matching strain, yielding 1980 additional monolayer calculations, and used these results as inputs for our EBA model analysis.
More calculation details can be found in the Method section and in Note S1 in the Supplementary Information (SI). All DFT outputs and relaxed crystal structures can be found in the attached json file in the SI.

\begin{figure*}[h]
\includegraphics[width=1.00\textwidth]{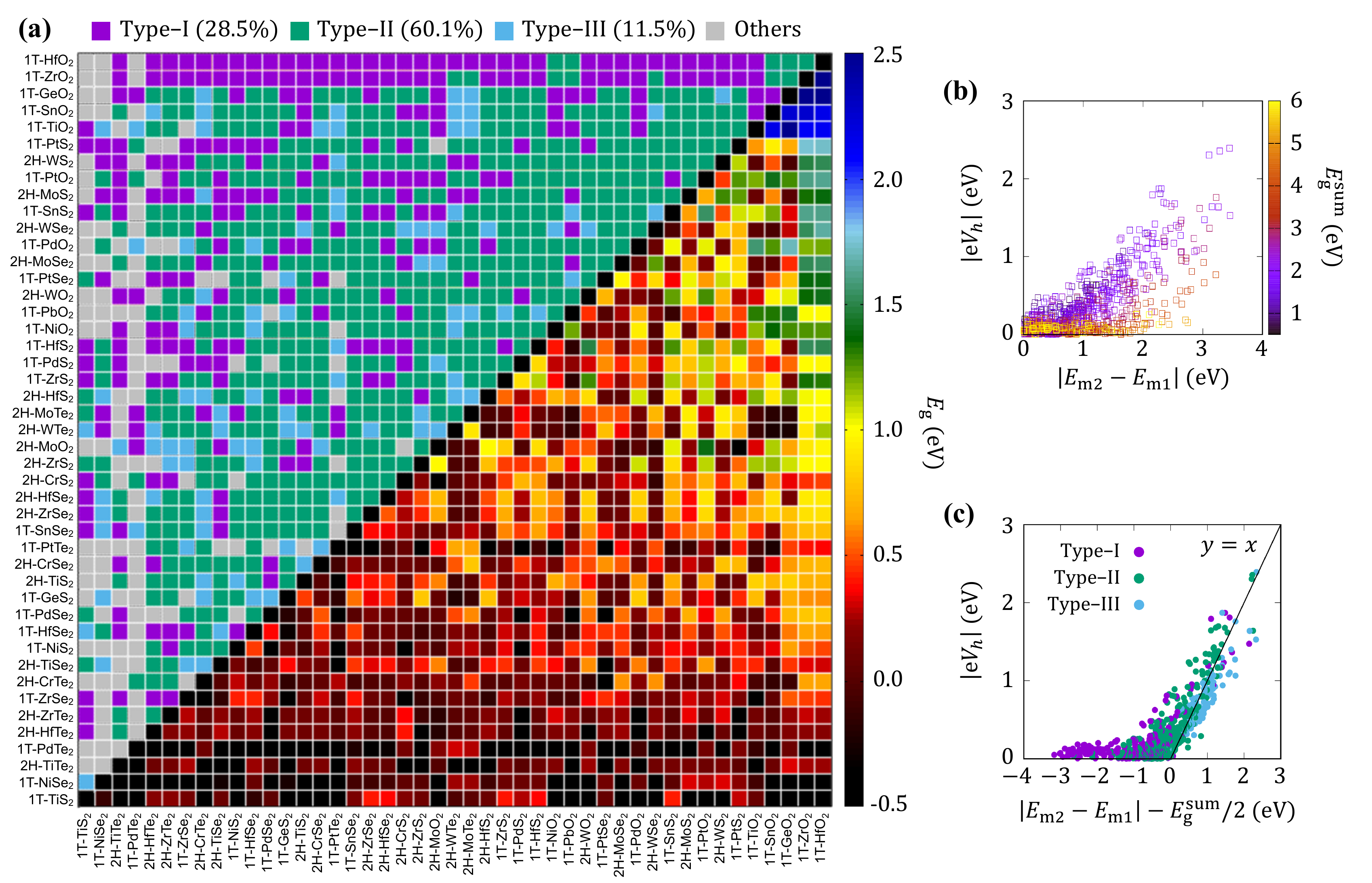}
\caption{
\textbf{High-throughput first-principles calculations for a total of 990 vdWHs}  
(a) Calculated bandgap and band alignment type of a total of 990 vdWHs. Electronic structures of TMD heterostructures calculated by PBE XC functional. Monolayer TMDs are ordered from left to right (bottom to top) according to the size of the bandgap. Top-left half and bottom-right half indicate the band alignment type and size of the bandgap of each heterostructures, respectively. 
(b) $|eV_{\rm{h}}|$ of vdWHs obtained from DFT calculations as function of the midgap difference ($|E_{\rm{m2}}-E_{\rm{m1}}|$) and a sum of the bandgaps of two monolayers ($E_{\rm{g}}^{\rm{sum}}$, color coordinates).
(c) Relation between $|eV_{\rm{h}}|$ and $|E_{\rm{m2}}-E_{\rm{m1}}|-E_{\rm{g}}^{\rm{sum}}/2$ of vdWHs. Color coordinates indicate the band alignment type obtained from DFT calculations.
}\label{fig2}
\end{figure*}

The calculated band alignment type (upper-left) and bandgap ($E_{\rm{g}}$, lower-right) of a total of 990 vdWHs were summarized in Fig.~\ref{fig2}(a).
The TMD monolayers are ordered from left to right (bottom to top) by their bandgap.
The calculated bandgap of vdHWs range from -0.32 to 4.33~eV, encompassing the full spectrum of possible vdWH configurations.
We found that the most common type of band alignment is type-II (60.1\%), which accounts for more than half of the total vdHWs population.
Nearly a quarter of vdWHs have a type-I band alignment. The type-I configuration dominates when the bandgap difference between two isolated layers is large (left or top), which is a reasonable expectation.
Finally, only 10$\%$ of the vdWHs are classified into as a type-III band alignment, a percentage that is substantially lower than previous predictions from the Anderson model (28.8\% of type-III vdWHs).~\cite{choudhary2023efficient}
This discrepancy suggests that the Anderson model overestimates the population of type-III vdWHs by underestimating the bandgaps because Anderson model does not account for charge exchange.
A small subset labelled “other” corresponds to systems that become metallic under the imposed strain/twist. (Note S2 in SI)
Since these violate the rigid-shift picture we have excluded them from this analysis, along with any bilayer in which either monolayer’s gap falls below 0.05 eV, leaving 847 stacks for quantitative analysis.

To further interrogate our first-principles dataset, we plotted $eV_{\rm{h}}$ of the vdWHs against the midgap difference ($|E_{\rm{m2}}-E_{\rm{m1}}|$) and the sum of the bandgaps of the two monolayers ($E_{\rm{g}}^{\rm{sum}}$), as shown in Fig.~\ref{fig2}(b).
According to the sLR (Eq.~(\ref{eq5})), $eV_{\rm{h}}$ is linearly proportional to $|E_{\rm{m2}}-E_{\rm{m1}}|$ and independent of $E_{\rm{g}}^{\rm{sum}}$. However, Fig.~\ref{fig2}(b) clearly shows that $eV_{\rm{h}}$ anti-correlates with $E_{\rm{g}}^{\rm{sum}}$.
The charge spillage at the interface provides an intuitive understanding for the bandgap‑dependence of EBA.
In general, if any vdWH has type-III band alignment type, a direct charge transfer occurs from the valence band of one layer to the conduction band of the other layer, establishing an interfacial built-in potential.
The magnitude of this potential grows with the broken gap, providing an intuitive explanation for the observed bandgap dependence of the EBA.
Such a charge transfer pathway will be absent in type-I and type-II band alignments.

To isolate the impact of charge spillage, we plotted $eV_{\rm{h}}$ as a function of $|E_{\rm{m2}} - E_{\rm{m1}}| - E_{\rm{g}}^{\rm{vac}}/2$ which represents $E^{\rm{broken}}$ within the Anderson model, as shown in Fig.~\ref{fig2}(c). (The Anderson model provides the band lineup before electrical contact)
The data reveal two distinct regimes depending on the sign of $|E_{\rm{m2}} - E_{\rm{m1}}| - E_{\rm{g}}^{\rm{vac}}/2$. For values below zero, corresponding to type-I and type-II alignments, $eV_{\rm{h}}$ shows no clear dependence, indicating no charge spillage. 
In contrast, for values above zero, $eV_{\rm{h}}$ increases roughly linearly with $|E_{\rm{m2}} - E_{\rm{m1}}| - E_{\rm{g}}^{\rm{vac}}/2$, indicating significant charge transfer effects.
Notably, the condition $|E_{\rm{m2}} - E_{\rm{m1}}| - E_{\rm{g}}^{\rm{vac}}/2 = eV_{\rm{h}}$, highlighted by a black solid line in Fig.~\ref{fig2}(c), marks heterostructures whose effective gap closes. 
The region below this line, therefore, primarily represents type-III vdWHs. This assignment is consistent with band alignment types obtained from DFT calculations (color coordinates in Fig.~\ref{fig2}(c)), although some outliers are observed due to non-ideal effects such as atomic relaxation and band hybridization.
These observations clearly demonstrate that $eV_{\rm{h}}$ cannot be described by a simple linear relation with $|E_{\rm{m2}} - E_{\rm{m1}}|$, motivating the explicit inclusion of charge spillage corrections in the linear response model discussed next.

\subsection*{Generalized linear response model}

To develop a simple and physically intuitive EBA model that incorporates the charge spillage contribution, we adopt two simplifying assumptions: (1) the density of states of valence and conduction bands of each monolayer are constant and are determined by its effective mass via $|m^{*}|/\pi\hbar^2$, and (2) the net charge density in each layer is spatially localized at the surface of the monolayer.
By solving the electrostatic Poisson equation under these approximations, we found that the interlayer dipole from the charge spillage ($eV^{{\delta}\rho}_{\rm{h}}$) is proportional to $E^{\rm{broken}}$ expressed as follows, (See Note S3 in SI for details)

\begin{equation}
eV^{{\delta}\rho}_{\rm{h}}
=\pm{\gamma}E^{\rm{broken}}
=\pm{\gamma}(|E'_{\rm{m2}}-E'_{\rm{m1}}|-\frac{1}{2}E_{\rm{g}}^{\rm{sum}}), \text{if } {\pm} E_{\rm{m2}} \mp  E_{\rm{m1}} > 0.
\label{eq7}
\end{equation}
Here, ${\gamma}$ is a dimensionless coefficient defined as
\begin{equation}
{\gamma}=
\begin{dcases}
\frac{d_{\rm{int}}{e^2}}
{\varepsilon_0\pi\hbar^2}
\frac{2|m^{*}_{\rm{2v}}||m^{*}_{\rm{1c}}|}{|m^{*}_{\rm{2v}}|+|m^{*}_{\rm{1c}}|}, & \text{if } E_{\rm{m2}}>E_{\rm{m1}} \\
\frac{d_{\rm{int}}{e^2}}
{\varepsilon_0\pi\hbar^2}
\frac{2|m^{*}_{\rm{1v}}||m^{*}_{\rm{2c}}|}{|m^{*}_{\rm{1v}}|+|m^{*}_{\rm{2c}}|}, & \text{if } E_{\rm{m2}}<E_{\rm{m1}}, 
\end{dcases}
\label{eq8}
\end{equation}
where $e$, $d_{\rm{int}}$, $\varepsilon_0$, and $m^{*}$ denote the elementary charge, interlayer distance, dielectric constant of the medium between two layers, vacuum permittivity, and effective masses of valence band maximum (v) and conduction band minimum (c). 
$\gamma$ acts as a dimensionless 2D quantum-capacitance parameter.~\cite{luryi1988quantum} 
Then, by combining Eqs.~(\ref{eq3}) and~(\ref{eq7}), the expression of the total electrostatic dipole can be written as 
\begin{equation}
eV_{\rm{h}}=
{\alpha}_{\rm{eff}}(E'_{\rm{m2}}-E'_{\rm{m1}})
+({\gamma}(E'_{\rm{m2}}-E'_{\rm{m1}})\mp\frac{1}{2}{\gamma}E_{\rm{g}}^{\rm{sum}}){\Theta}(E^{\rm{broken}}), 
\text{if } {\pm}E_{\rm{m2}} {\mp} E_{\rm{m1}}>0,
\label{eq9}
\end{equation}
where ${\Theta}$ is a Heaviside step function because $eV^{{\delta}\rho}_{\rm{h}}$ must vanish in type-I and type-II vdWHs. 

Solving Eqs.~(\ref{eq1}) and~(\ref{eq9}) self-consistently, exactly as in the sLR model, yields the interfacial potential drop $eV_{\rm{h}}$.
During the iteration, the charge spillage term vanishes automatically if the stack is classified as type-I or type-II, so the result reverts to the sLR expression (Eq.~(\ref{eq4})).
Consequently, the spillage correction contributes only for type-III vdWHs, for which,
\begin{equation}
eV_{\rm{h}}=
\frac{{{\alpha}_{\rm{eff}}+\gamma}}{{{\alpha}_{\rm{eff}}+\gamma}+1}(E_{\rm{m2}}-E_{\rm{m1}})\mp\frac{\gamma}{2({\alpha}_{\rm{eff}}+\gamma+1)}E_{\rm{g}}^{\rm{sum}}, \text{if } {\pm} E_{\rm{m2}}{\mp}E_{\rm{m1}} > 0.
\label{eq10}
\end{equation}
In the limit of ${\gamma}{\rightarrow}\infty$, $eV_{\rm{h}}$ approches to $E^{\rm{broken}}$ indicating that type-III vdWH is not allowed (zero bandgap). In the opposite limit, i.e. the limit of ${\gamma}{\rightarrow}0$, Eq.~({\ref{eq10}}) is reduced to Eq.~(\ref{eq4}) -without the need for charge spillage corrections.
As such, Eq.~({\ref{eq10}}) can be viewed as a generalized model to the sLR.
For practical implementation of the gLR model, we have employed a hyperbolic tangent function, $\frac{1}{2}\tanh(E_{\rm{g}}^{\rm{broken}}/\eta) + \frac{1}{2}$, instead of ${\Theta}(E_{\rm{g}}^{\rm{broken}})$ to improve the numerical stability of the self-consistent calculation. 

\begin{figure*}[t]
\includegraphics[width=1.00\textwidth]{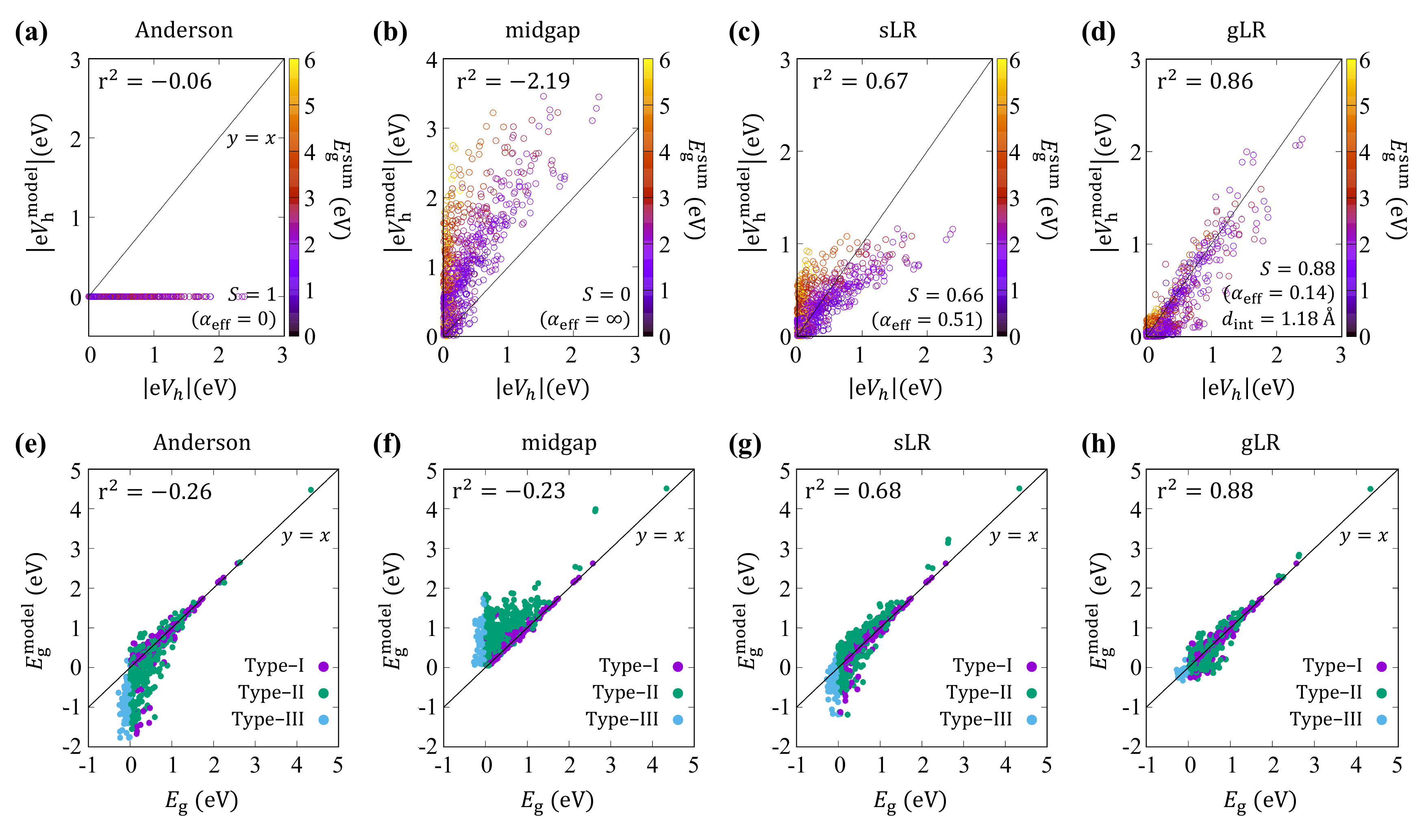}
\caption{
\textbf{Accuracy of generalized band alignment model}  
Comparison between $eV_{\rm{h}}$ obtained by DFT calculations and $eV_{\rm{h}}^{\rm{model}}$ predicted by the (a) Anderson, (b) midgap, (c) sLR and (d) gLR models, respectively.
Comparison between calculated and predicted EBA of vdWHs predicted by the (e) Anderson, (f) midgap, (g) sLR and (h) gLR models. 
$E_{\rm{g}}^{\rm{model}}$ and $E_{\rm{g}}$ indicate $E_{\rm{g}}$ of vdWHs predicted by the model and obtained by DFT calculations, respectively. The color coordinates indicate the band alignment type predicted by DFT calculation.
The performance of model is quantitatively evaluated using a coefficient of determination ($r^2$).
}\label{fig3}
\end{figure*}

To benchmark the gLR model, we compared its predictions of $eV_{\rm{h}}$ against other EBA models for all vdWHs, as shown in Fig.~\ref{fig3}(a-d). 
The Anderson and midgap models consistently overestimate or underestimate $eV_{\rm{h}}$ with poor correlation to the DFT results (negative $r^2$ values). By construction, the Anderson and midgap models fix $eV_{\rm{h}}^{\rm{model}}$ to be either 0 or $(E_{\rm{m2}}-E_{\rm{m1}})$, which always underestimates or overestimates $eV_{\rm{h}}$, respectively. 
The sLR  improves on these baselines by introducing only a single screening parameter, $S$ or ${\alpha}_{\rm{eff}}$.
In bulk semiconductor heterostructures, the heterojunction interfacial $S$ can be described as $\frac{1}{2}(\frac{1}{{\varepsilon}_1}+\frac{1}{{\varepsilon}_2})$~\cite{tersoff1984theory2,vazquez2005energy} where ${{\varepsilon}_1}$ and ${{\varepsilon}_2}$ indicate dielectric constants of two semiconducting materials. 
However, ${{\varepsilon}}$ is not precisely defined in 2D materials, because it requires a layer thickness that is difficult to define in the 2D limit.~\cite{tian2019electronic}
Therefore, as a first attempt, we employed a constant value of $S$ across for all vdWHs.~\cite{PhysRevB.103.205129}
Treating $S$ as a global fitting parameter, $S$ was determined to be 0.66 with $r^2=0.67$.
Although the sLR model shows much better performance compared to the Anderson and midgap models, it still underestimates (overestimates) $eV_{\rm{h}}$ when $eV_{\rm{h}}$ is relatively large (small). 
This is precisely the regime where charge spillage across the vdW gap (neglected in sLR) becomes critical, especially for broken gap (type-III) stacks. 

The quantum capacitance $\gamma$ quantifies the relationship between the broken gap and the charge-spillage-induced interfacial dipole.
In Eq.~(\ref{eq8}), $\gamma$ contains a unknown parameter, $d_{\rm{int}}$, which cannot be inferred from isolated-layer properties.
Therefore, $d_{\rm{int}}$ is used as a global constant and fitting it for all vdWHs, together with the screening parameter $S$.
$S$ and $d_{\rm{int}}$ were determined to be 0.88 and 1.18~{\AA}, respectively.
Using these two constants, the gLR model predicts $eV_{\rm{h}}$ with a very high accuracy ($r^2=0.86$) as shown in Fig.~\ref{fig3}(d), across the full spectrum of band alignments from type-I to broken gap type-III. The result underscores the model’s power as a lightweight yet physically grounded surrogate for exploring the enormous design space of vdW heterostructures.

For a more comprehensive evaluation of EBA models, we compare the predicted bandgaps of vdWHs against DFT results, as shown in Fig.~\ref{fig3}(e-h).
The DFT benchmark values are colour-coded by the true alignment type to highlight systematic errors.
As expected, the Anderson rule performs tolerably for wide-gap insulators but totally collapses for narrow-gap stacks because it neglects any interlayer interaction or charge transfer. 
The midgap rule has an error in the opposite direction and marginally describes the bandgap of type-I cases.
Introducing a single screening factor already lets the sLR model outperform both baselines.
However, as discussed above, it fails for large bandgap vdWHs and also for type-III vdHWs.
By contrast, the gLR model, which augments sLR with the quantum capacitance term, achieves the highest fidelity across all alignment types with a $r^2=0.88$.

At this point, it is worth discussing the transferability of the fitting parameters used in this study.
In our previous research, we found an optimal value of $S$=0.66 (or $\alpha_{\rm{eff}}=0.52$) for group IV monochalcogenides,~\cite{PhysRevB.103.205129} which is smaller than the value used for TMDs ($S$=0.88 (or $\alpha_{\rm{eff}}=0.14$)). 
Consequently, the parameters $S$ found in this study are specific to TMDs and not universal. 
Intuitively, the value of $S$ might be affected by many physical properties of monolayers such as dielectric function, 2D polarizability,~\cite{tian2019electronic} orbital characteristics, and related factors.
A systematic mapping of $S$ across diverse 2D chemistries, and a search for unifying trends, lie beyond the present scope but will be important directions for future work.


\subsection*{Validation of generalized linear response model using machine learning analysis}

\begin{figure*}
\includegraphics[width=1.00\textwidth]{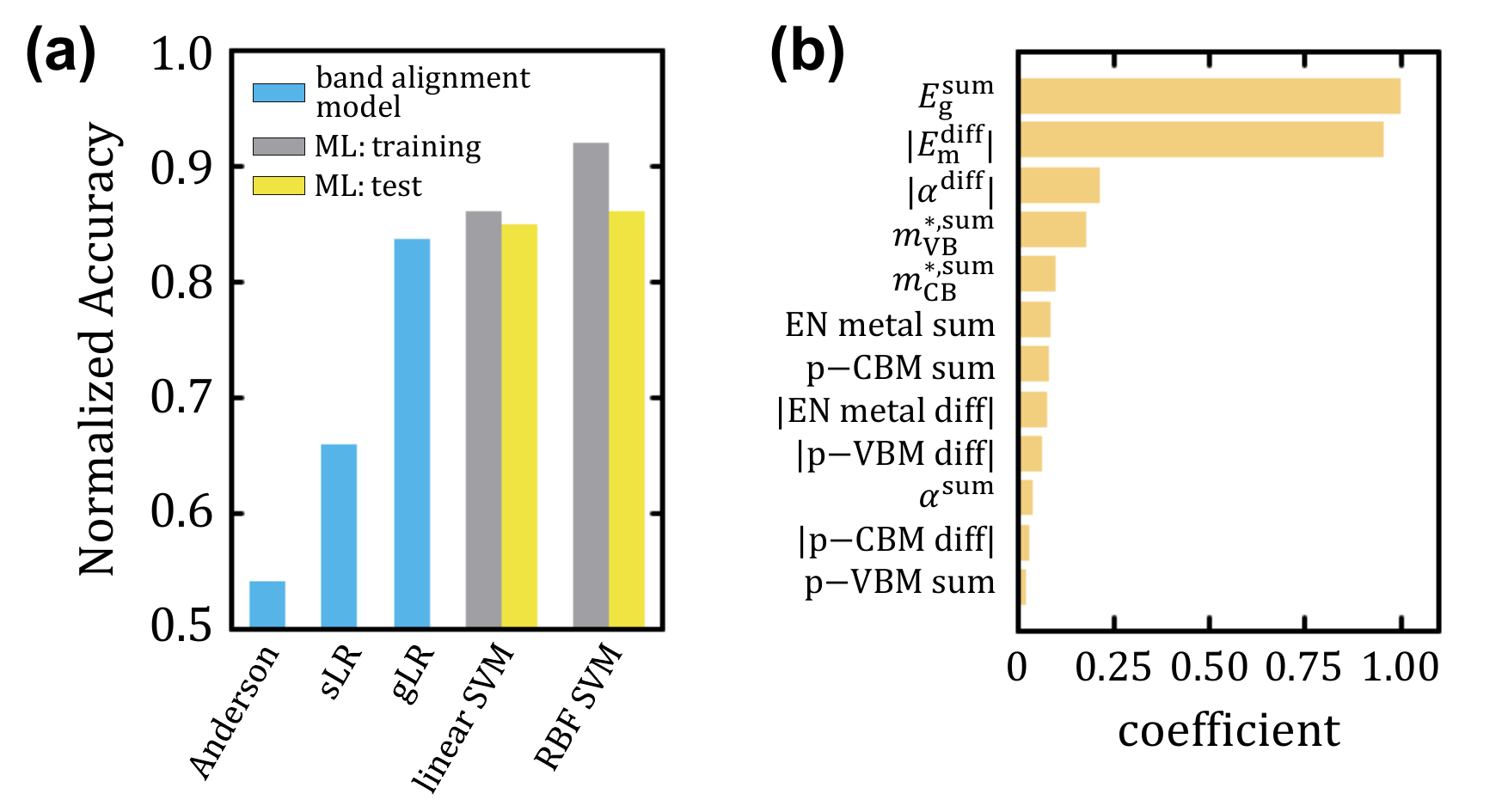}
\caption{
\textbf{Machine learning validation on the generalized band alignment model}
(a) The performance metric score of normalized accuracy (NA)
for the Anderson, sLR, gLR models, and a linear and Radial Basis Function (RBF) Support Vector Machine (SVM) models, respectively. 
(b) Input features importance extracted from the linear SVM model.
}
\label{fig4}
\end{figure*}

The remarkable accuracy of the gLR model naturally raises fundamental and important questions. 
Can we further improve the accuracy of the gLR model? 
What other physical input features also contribute to the EBA?
Although, in principle, many other physical properties of 2D monolayer can contribute to the EBA, it is non-trivial to measure the relative importance among various physical properties in classical physics-based analysis. 
To address these questions, we employed machine learning (ML) modeling, which involves identifying the relationship between the selected input features and a target output based on available data.~\cite{cherkassky_learning_2007, tan_introduction_2006}
For the ML modeling, we designed a binary classification problem on whether a particular vdWH exhibits small band offset (SBO) or not.~\cite{choudhary2023efficient, willhelm2022predicting,lee_methodological_2022}.
Here, a vdWH is classified into SBO when it satisfies two criteria: (1) it has either a type-II or type-III band alignment, and (2) its absolute bandgap is less than $0.15$~eV. 
Materials that satisfy these criteria are labeled as ``positive'' class, while those that do not are classified as ``negative''. 
Note that vdWHs exhibiting SBO are particularly interesting because their band alignment type can be tuned via an external gate voltage, owing to the sufficiently small $|E_{\rm{g}}|$.~\cite{Fan2019,Lei2019,iordanidou2022electric} 

Support Vector Machine (SVM) classifiers \cite{cortes_support-vector_1995, cherkassky_learning_2007, tan_introduction_2006} with linear and Radial Basis Function (RBF) kernels were chosen to classify SBO vdWHs. 
Selected input features include not only the key features used in the gLR model, but also the orbital characteristics of the valance and conduction bands, which are relevant to band hybridization. 
Furthermore, the electronegativity of metallic atom of monolayer has been included as an input feature. 
In order to impose invariance with respect to the stacking order of monolayers within a heterostructure, the input features from each monolayer in heterostructure were combined using absolute difference (denoted as $|\text{diff}|$) and summation (sum) operations. 
A complete list of input features and the data-preprocessing procedure for eliminating correlated features are detailed in Note S4 in SI. 
The final dataset for machine learning modeling consists of 847 vdWHs with 12 input features. Out of these 847 vdWHs, 257 (30.1\%) of them have positive class and 590 (69.9\%) have negative class. 

Since available data is unbalanced, prediction accuracy of ML model is estimated using normalized accuracy (NA)~\cite{lee_methodological_2022, tan_introduction_2006}.
NA measures the model prediction accuracy adjusted for class imbalance, so that NA=0.5 corresponds to random guessing and NA=1.0 to perfect prediction. 
See Machine Learning Modeling in Note S5 in SI for detailed explanation of the modeling procedure and performance metrics. 
Figure~\ref{fig4}(a) shows the NA for the band alignment and SVM models under the same classification setting. 
These results show that SVM models significantly outperformed the Anderson and sLR models. 
Surprisingly, the gLR shows performance comparable to SVM models, indicating remarkable accuracy of the gLR. 
The RBF SVM model performs the best among all models, suggesting that incorporating non-linearity can further improve model performance.

Figure~\ref{fig4}(b) illustrates the feature importance extracted from the linear SVM model~\cite{cherkassky_learning_2007}. 
Notably, the two most important features are the bandgap sum $E^{\rm{sum}}_{\rm{g}}$ and the midgap difference $|E^{\rm{diff}}_{\rm{m}}|$, where both exhibit similar relative importance. 
These are followed by the 2D polarizability difference ($|\alpha^{\rm{diff}}|$), and the effective masses ($m^{*,\rm{sum}}_{\rm{VB}}$ and $m^{*,\rm{sum}}_{\rm{CB}}$), in descending order of importance.
These results explain why classical models, such as the Anderson and sLR models, fall short of predicting vdWH SBO, as they neglect the critical role of the monolayer bandgap feature. 
Furthermore, features employed in the gLR exhibit greater importance compared to other features.
Therefore, we conclude that that a physically motivated gLR model is very close to being an optimal approach.
Encoding physical principles into ML framework can enhance the prediction performance of EBA and reveal the relationship between the intrinsic properties of monolayers and their heterostructure. 
Further analysis of ML models is presented in Note S6 in SI.

\section*{Conclusion}

We have developed a generalized linear response (gLR) model that retains the simplicity of classical band alignment rules while reaching useful accuracy for 2D vdW systems. Our calculations show that traditional Anderson/midgap approaches, which rely solely on the CNL offset, break down for broken gap (type-III) stacks because they ignore the dipole created by charge spillage across the vdW gap. Embedding this missing dipole, quantified through a quantum capacitance term, into a linear response framework raises the predictive $r^2{\approx}0.9$ across all alignment types. Feature importance metrics from machine learning confirm that the two descriptors used in gLR; the CNL difference and the sum of the isolated-layer gaps, indeed dominate the underlying physics, suggesting the model is close to minimal and therefore broadly transferable. The gLR scheme thus offers an efficient, physically transparent surrogate for high-throughput screening of the vast vdWH design space and a deeper window into the electrostatics governing 2D heterostructures. The insights developed from this study could help speed the discovery and implementation of novel heterostructure devices based upon 2D materials. 


\section*{Method}

To systematically investigate electronic structures of various 2D materials and their heterostructures, we performed the first-principles calculations based on density functional theory~\cite{Kohn1965} as implemented in Vienna \textit{ab initio} simulation package (VASP)~\cite{Kresse1996}. 
The projector-augmented wave pseudopotentials~\cite{{Blochl1994},{Kresse1999}} and the generalized gradient approximation of Perdew-Burke-Ernzerhof (PBE)~\cite{Perdew1996} exchange-correlation (XC) functional were used 
The kinetic energy cutoff for the plane wave basis was chosen to be 400~eV.
The Brillouin zone was sampled by 21$\times$21$\times$1 k-mesh grid for the primitive unitcell of TMD monolayers.

The heterostructures were generated by the coincidence lattice method~\cite{PhysRevB.71.235415,Wang2015_JPCC} to reduce the artificial strain due to the lattice mismatch.
To avoid very large unitcell, we allowed a maximum 5~\% of strain, and the average value of strain for all heterostructures was 2.0~\%.
The Brillouin zone of heterostructures was sampled by a proportionally reduced k-mesh grid compared to the TMD monolayers.
To mimic 2D layered structure in periodic cells, we included a sufficiently large vacuum region ($c=30$~{\AA}) between neighboring cells along the out-of-plane direction, and the van der Waals interaction was treated as Grimme-D3 method.~\cite{grimme-d3}

\section*{Data availability}
All data generated and/or analyzed during this study are included in this article, its Supplementary Information file, and the corresponding website: \url{https://pages.github.umn.edu/tlow/Seungjun_Lee-Band_Alignment_Model/}. 
All raw data used in the current study are available from the corresponding author under request.

\acknowledgments
This work is supported by NSF DMREF-1921629 and in part by NSF ECCS-1542202.
S. L. is also supported by Basic Science Research Program through the National Research Foundation of Korea funded by the Ministry of Education (NRF-2021R1A6A3A14038837).



\end{document}